\documentclass[12pt]{iopart}

\newcommand{\nK}{K$^0_{\mathrm S}\ $}
\newcommand{\nL}{$\Lambda$\ }
\usepackage{graphicx}  
\begin{document}

\title[\nK and \nL production in Pb--Pb collisions with the ALICE
  experiment]{\nK and \nL production in Pb--Pb collisions with the ALICE experiment}

\author{Iouri Belikov for the ALICE Collaboration}

\address{IPHC, Universit\'{e} de Strasbourg, CNRS-IN2P3, 23 rue du Loess,
  BP28, 67037 Strasbourg cedex 2 }
\ead{Iouri.Belikov@in2p3.fr}

\begin{abstract}
We present the study of \nK and \nL production performed with the ALICE 
experiment at the LHC in Pb--Pb collisions at $\sqrt{s_\mathrm{NN}}=2.76$~TeV
and pp collisions at $\sqrt{s}=0.9$ and 7~TeV. The \nK and \nL particles 
are reconstructed via their V0 decay topology allowing their identification 
up to high transverse momenta. The corresponding baryon/meson ratios as a 
function of transverse momentum are extracted for Pb--Pb collisions in 
centrality bins and in the transverse momentum range from 1 to 6~GeV/$c$. 
They are also compared with those measured in pp events at the LHC
energies of 0.9 and 7~TeV as well as in Au--Au collisions at 
$\sqrt{s_\mathrm{NN}} = 62.4$ and 200~GeV from RHIC.
\end{abstract}



\section{Introduction}
One of the most interesting results obtained at RHIC in Au--Au collisions
was the observation that the baryon (anti-baryon) production
at intermediate transverse momenta becomes comparable to that of 
mesons~\cite{Adams:2006wk, Adcox:2003nr}.
The topological
decay reconstruction of \nK and \nL provides a unique opportunity
to extend the baryon and meson identification to much larger transverse
momenta than would be possible using conventional particle
identification methods. 
Measurement performed by the STAR
Collaboration (see, for example,~\cite{Adams:2006wk, Lamont:2007ce}), 
showed
that the baryon/meson ratio reaches its maximum at $p_\mathrm{T}\sim
2.5$~GeV/$c$ and starts decreasing at higher momenta. The maximum value
of the $\Lambda$/\nK ratio in central collisions was found to exceed
unity.

The question of why, in nucleus-nucleus collisions, baryons at
intermediate $p_\mathrm{T}$ appear to be more easily produced than mesons
is still open. Possible explanations involve
interplays between soft and hard mechanisms of particle production (also at
the partonic level) like those discussed, for example, in
Ref.~\cite{Fries:2003fr}.  
The evolution of the baryon/meson ratio with
collision energy may yield additional information about this ``baryon
anomaly''. 

In this article, we present the $\Lambda$/\nK ratios measured by the ALICE
experiment at the LHC in 1.1$\times$10$^7$ minimum bias Pb--Pb events 
at $\sqrt{s_\mathrm{NN}} = 2.76$~TeV as a function of transverse momentum 
and for different collision centrality bins, as well as in pp    
collisions at $\sqrt{s}=0.9$ and 7~TeV.
 
\section{Reconstruction of the \nK and \nL in ALICE}
The ALICE experiment is well suited for \nK and \nL reconstruction
over a wide momentum range.  For the results discussed here, the momentum
range $1 < p_\mathrm{T} < 6$~GeV/$c$ is defined by our current level of
understanding the systematic uncertainties.

The \nK and \nL particles were reconstructed via their V0 decay 
topology~\cite{Aamodt:2011zz}. The method was the same for both pp and Pb--Pb
collisions. 
The typical reconstruction efficiencies (extracted from 
Monte Carlo studies) were about 40~\% for \nK and 30~\% for \nL 
at $p_\mathrm{T}\sim 3$~GeV/$c$ (close to the $\Lambda$/\nK maximum). 
In the momentum range $2.5 < p_\mathrm{T} < 5.5$~GeV/$c$, the variation of
the ratio of the reconstruction efficiencies for \nK and \nL was 
no larger than 1--2~\%.  

The $p_\mathrm{T}$ spectra of \nK 
obtained in centrality bins in Pb--Pb collisions
were compared with the spectra of charged kaons reconstructed by the
ALICE Time Projection Chamber and the Time Of Flight detector.
At $p_\mathrm{T} > 1$~GeV/$c$, the two sets of spectra agreed within 
1--2~\%~\cite{alex}.  
    
The spectra of \nL were corrected for the contribution of $\Lambda$'s
coming from decays of $\Xi^-$ and $\Xi^0$. This was done by re-scaling the
corresponding distributions extracted from Monte Carlo simulations
with the $p_\mathrm{T}$ spectrum of $\Xi^-$ reconstructed in real data,
assuming that the number $\Xi^0$ is proportional to the reconstructed number
of $\Xi^-$.
The obtained feed-down corrections turned out to be of the order of 20~\%
and changed within only a few per cent as a function of event
centrality  and transverse momentum.
 
The efficiency and feed-down corrections
were checked with the life-time distributions for the V0 particles.
These distributions were corrected as functions of two variables,
$p_\mathrm{T}$ and decay length. This was done for all the
event centrality bins separately. The statistical error of the reconstructed
life times for \nK and \nL was of the order of
1~\%.  However, the systematic deviation from the corresponding nominal values
was  $\sim 3$--4~\% in the case of the most central events.  

Altogether, we considered the following main sources of systematic
uncertainties,
listed here along with their contributions to the overall uncertainty:
signal extraction (3~\%),
efficiency correction (7~\% for $p_\mathrm{T}<1$~GeV/$c$, 1~\% for
 $p_\mathrm{T} > 2.5$~GeV/$c$),
feed-down correction (5~\%),
admixture of $\Lambda$'s generated in the detector material (2~\%).

\section{Preliminary results}

\begin{figure}
\centering
\includegraphics[width=0.47\textwidth, height=0.37\textwidth]{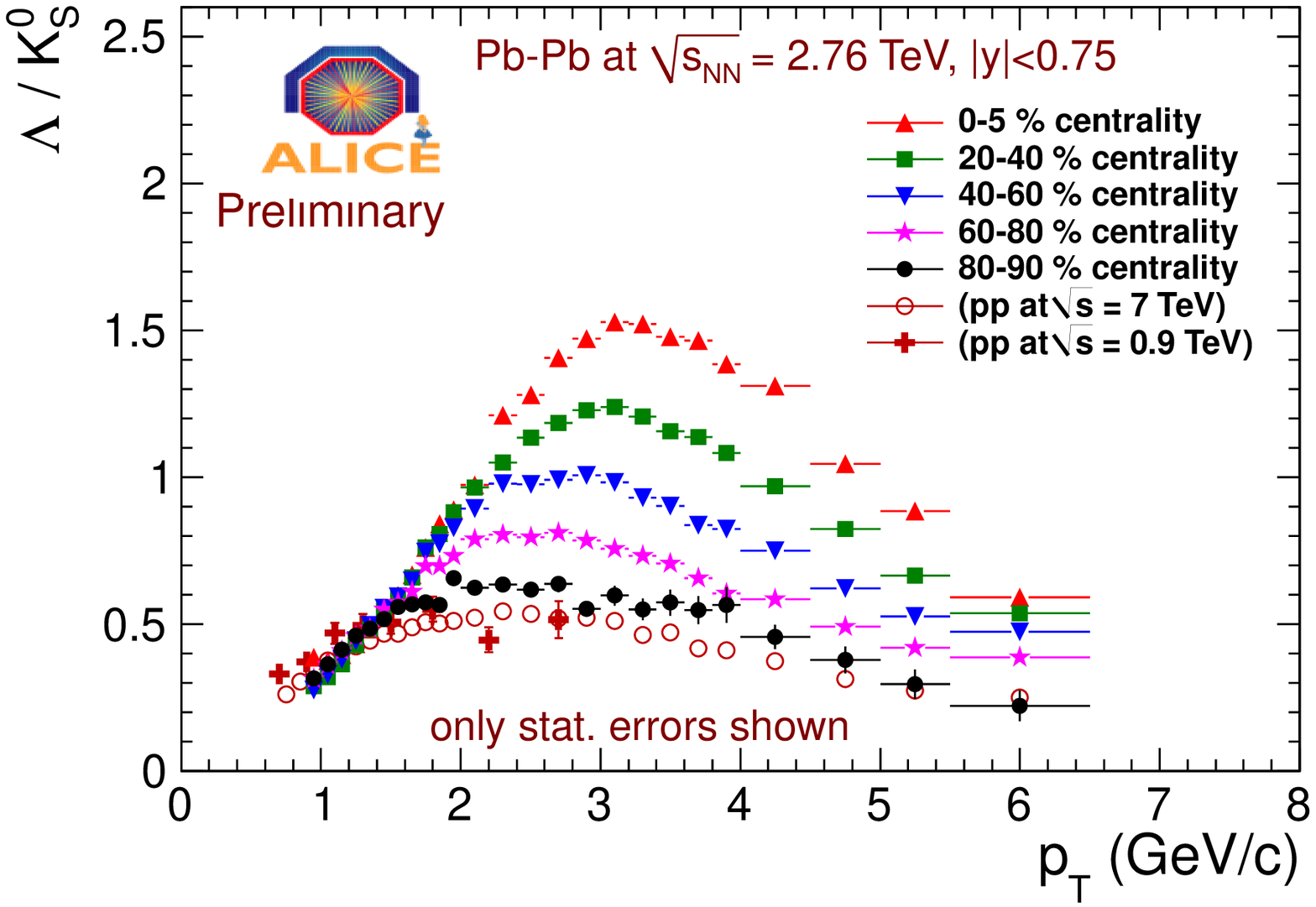}
\includegraphics[width=0.52\textwidth, height=0.35\textwidth]{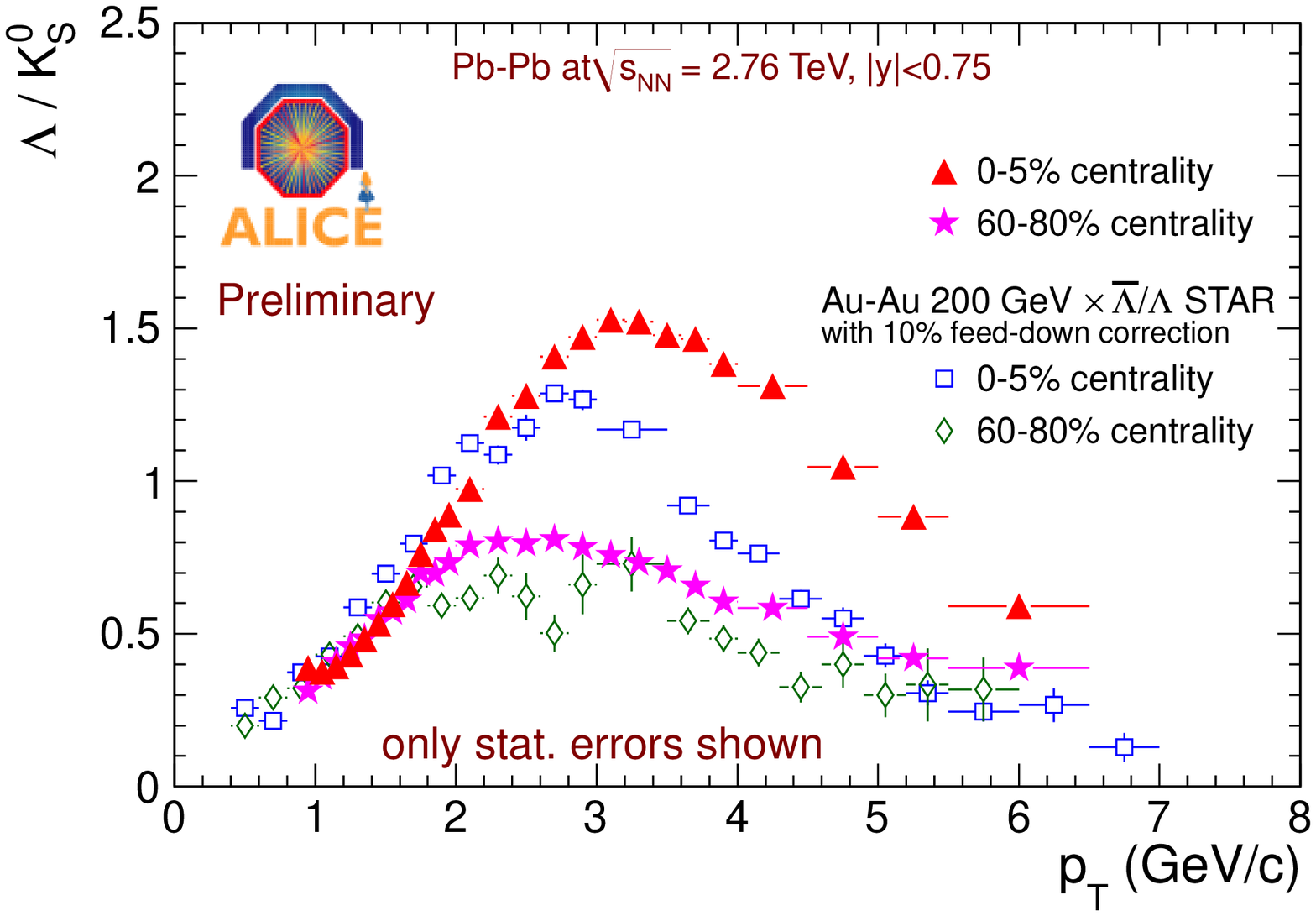}

\vskip -2 mm

\caption{Left: $\Lambda$/\nK ratios as a function of $p_\mathrm{T}$ for
  different centralities in Pb--Pb collisions at 
$\sqrt{s_\mathrm{NN}} = 2.76$~TeV, and also for minimum bias pp collisions
  at $\sqrt{s} = 0.9$ and 7~TeV. Right: selected $\Lambda$/\nK ratios shown 
  on the left compared with those measured in Au--Au collisions at 
  $\sqrt{s_\mathrm{NN}} = 200$~GeV.
\label{fig1}
}
\end{figure}

The $\Lambda$/\nK ratios as a function of $p_\mathrm{T}$ for
different centralities in Pb--Pb collisions at 
$\sqrt{s_\mathrm{NN}} = 2.76$~TeV are shown in Fig.~\ref{fig1} (left). 
The same ratios for minimum bias pp events at 0.9 and 7~TeV are also given. 
The baryon/meson ratio in  pp interactions 
always stays below~1 and  
is quite similar to what is observed in peripheral Pb--Pb
collisions. As the collision centrality increases,
the baryon/meson ratio 
develops a maximum at
$p_\mathrm{T}\sim 3$~GeV/$c$ reaching a value of $\sim 1.5$ for the 0--5~\% 
most central events.

\section{Comparison with previous measurements}

Comparing these preliminary $\Lambda$/\nK ratios with those measured by the
STAR Collaboration in Au--Au collisions at
$\sqrt{s_\mathrm{NN}} = 200$~GeV, we notice that, in
the case of most central events, the baryon/meson ratio at the LHC
decreases less rapidly with $p_\mathrm{T}$ than at RHIC~(see
Fig.~\ref{fig1}~(right)).  The preliminary STAR data 
points~\cite{Lamont:2007ce} shown in this figure are multiplied by 
the $\bar{\Lambda}/\Lambda = 0.8$ factor calculated from the data 
reported in~\cite{Adams:2006ke}
(to account for the non-unity of the anti-baryon/baryon ratio at RHIC)   
and subtracted a 10~\% feed-down correction quoted in~~\cite{Adams:2006wk}.

\begin{figure}
\centering
\includegraphics[width=0.48\textwidth,height=0.36\textwidth]{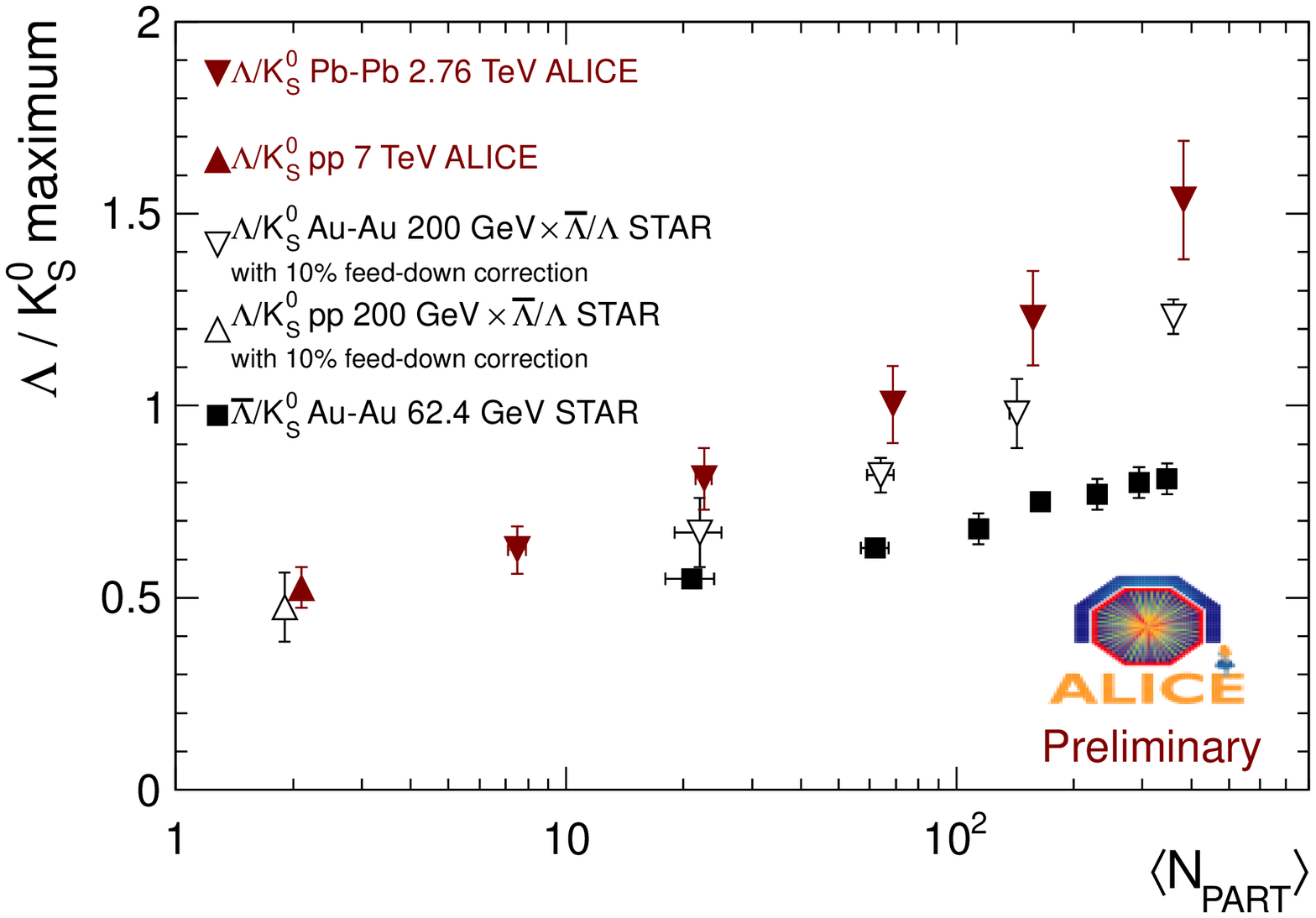}
\includegraphics[width=0.48\textwidth,height=0.36\textwidth]{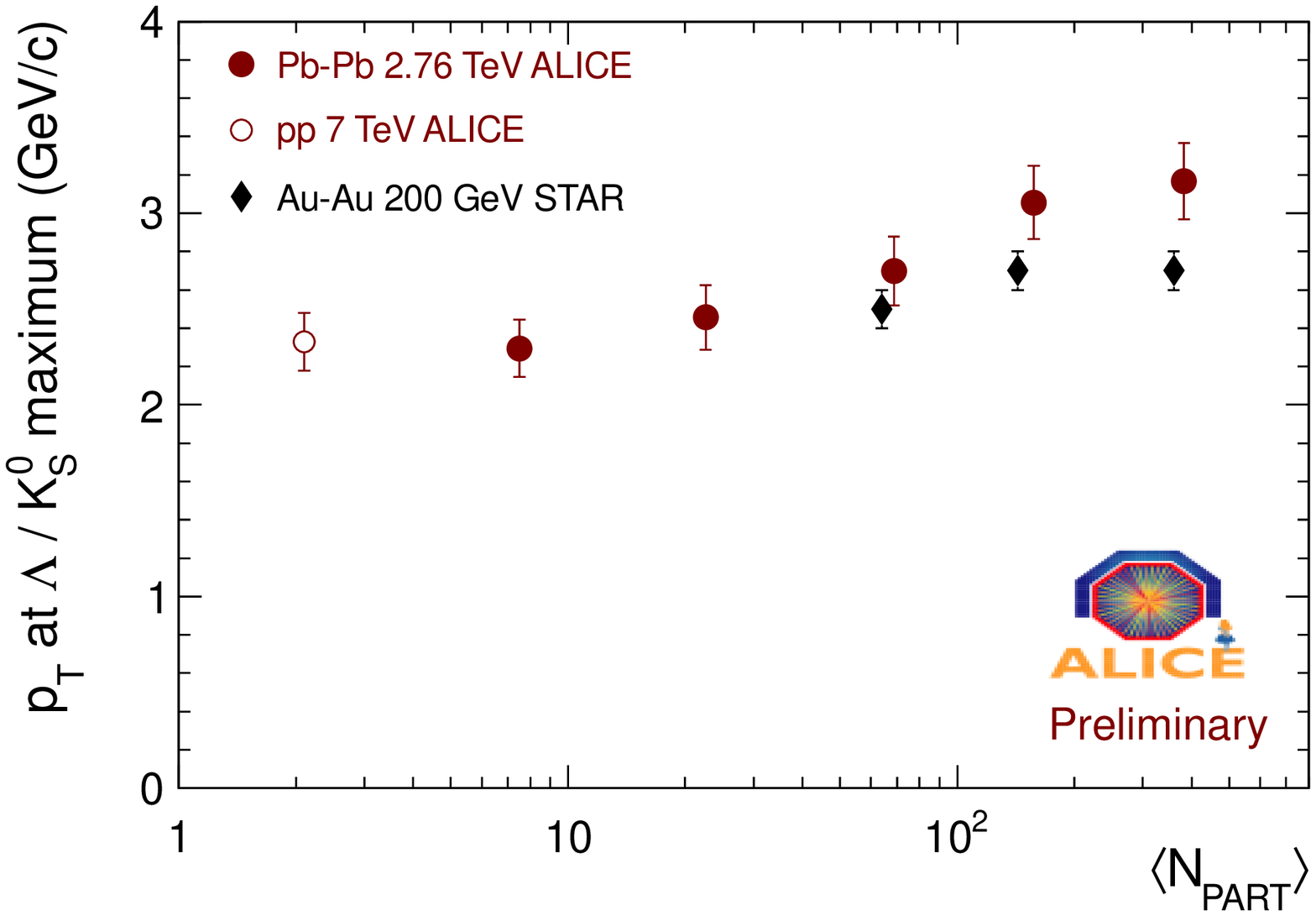}
\caption{Left: maximum value of the $\Lambda$/\nK ratio as a function
of number of participants compared between different colliding systems and
energies.
Right: position in $p_\mathrm{T}$ of the $\Lambda$/\nK maximum as a
  function of number of participants compared between Pb--Pb at
  $\sqrt{s_\mathrm{NN}} = 2.76$~TeV, pp at $\sqrt{s} = 7$~TeV
 and Au--Au at $\sqrt{s_\mathrm{NN}} =
  200$~GeV collisions.
\label{fig2}
}
\end{figure}

A comparison between the maximum values of (anti-)baryon/meson 
ratios measured by ALICE in Pb--Pb collisions at 
$\sqrt{s_\mathrm{NN}} = 2.76$~TeV  and those obtained by STAR in Au--Au
events at $\sqrt{s_\mathrm{NN}} = 62.4$ and 200~GeV is presented in
Fig.~\ref{fig2} (left). 
The Au--Au points at 62.4~GeV are plotted as they are published 
in~\cite{Aggarwal:2010ig}. 
To compare the STAR measurements at 200~GeV with the ALICE results,
we multiply the STAR values
from~\cite{Lamont:2004qy} by the same $\bar{\Lambda}/\Lambda$
factor 
and apply the same feed-down correction as mentioned above.
As is evident in Fig.~\ref{fig2} (left), the maximum value of the $\Lambda$/\nK ratio
increases with the beam energy.

The position in $p_\mathrm{T}$ of the $\Lambda$/\nK maximum 
measured at Pb--Pb collisions at $\sqrt{s_\mathrm{NN}} = 2.76$~TeV
is slightly shifted towards higher transverse momenta 
with respect to that observed in Au--Au events at
$\sqrt{s_\mathrm{NN}} = 200$~GeV, as shown in the Fig.~\ref{fig2} (right).  

\section{Conclusions}
We have presented the measurements of
$\Lambda$/\nK ratios in Pb--Pb collisions at $\sqrt{s_\mathrm{NN}} = 2.76$~TeV
performed with the ALICE experiment at the LHC. 
The ratios are 
compared to those measured by ALICE in pp collisions at $\sqrt{s}
= 0.9$ and 7~TeV as well as with the STAR results 
in Au--Au events at $\sqrt{s_\mathrm{NN}} = 62.4$ and 200~GeV.

In Pb--Pb collisions, 
the $\Lambda$/\nK ratio as a
function
of the transverse momentum shows a broad maximum around 
$p_\mathrm{T}\sim 3$~GeV/$c$. The maximum value of the ratio 
increases with the
collision centrality reaching the value of $\sim 1.5$ for the 0-5~\% 
most central events. 

As the collision centrality decreases, the maximum of the $\Lambda$/\nK
ratio becomes less pronounced and diminishes to a value of $\sim 0.6$. The same
behaviour of the ratio is observed in pp events at $\sqrt{s} = 0.9$ and
7~TeV which bracket in energy the Pb--Pb results reported here.

Comparison with similar measurements performed by the STAR Collaboration
in Au--Au collisions at $\sqrt{s_\mathrm{NN}} = 62.4$ and 200~GeV shows
that the value at the $\Lambda$/\nK maximum increases with the beam energy.
At the same time, the position of the maximum in $p_\mathrm{T}$ shifts
towards higher transverse momenta. The magnitude of this shift is 
smaller than it was predicted, for example, in Ref.~\cite{Fries:2003fr}. 
However, the baryon enhancement in central
nucleus-nucleus collisions at the LHC decreases less rapidly with 
$p_\mathrm{T}$ and, at $p_\mathrm{T}\sim 6$~GeV/$c$, is
a factor $\sim 2$ higher compared with that at RHIC.

\end{document}